\documentclass[twocolumn,10pt]{IEEEtran}
\usepackage{ifpdf, flushend}

\ifCLASSINFOpdf
  \usepackage[pdftex]{graphicx}
  \graphicspath{{../pdf/}{../jpeg/}}
  \DeclareGraphicsExtensions{.pdf,.jpeg,.png}
\else
  \usepackage[dvips]{graphicx}
  \graphicspath{{../eps/}}
  \DeclareGraphicsExtensions{.eps}
\fi
\usepackage[cmex10]{amsmath}
\usepackage {amssymb}
\usepackage{algorithmic}
\usepackage{array}
\usepackage{mdwmath}
\usepackage{mdwtab}
\usepackage{eqparbox}
\usepackage{url}
\usepackage{hyperref}
\usepackage{algorithm}
\usepackage{algorithmic}
\usepackage{epstopdf,cite,color}
\usepackage{amsthm} 

\renewcommand{\baselinestretch}{1}

\newcommand{\beq}{\begin{equation}}
\newcommand{\eeq}{\end{equation}}
\newcommand{\beqn}{\begin{eqnarray}}
\newcommand{\eeqn}{\end{eqnarray}}
\newcommand{\beqno}{\begin{eqnarray*}}
\newcommand{\eeqno}{\end{eqnarray*}}
\newcommand{\bma}{\begin{displaymath}}
\newcommand{\ema}{\end{displaymath}}
\newcommand{\bnu}{\begin{enumerate}}
\newcommand{\enu}{\end{enumerate}}
\newcommand{\bce}{\begin{center}}
\newcommand{\ece}{\end{center}}
\newcommand{\btb}{\begin{tabular}}
\newcommand{\etb}{\end{tabular}}

\begin{document}

\title{Privacy-Aware Framework of Robust Malware Detection in Indoor Robots: Hybrid Quantum Computing and Deep Neural Networks}
\author{Tan~Le,~\IEEEmembership{Member,~IEEE,} Van~Le, and Sachin~Shetty,~\IEEEmembership{Senior Member,~IEEE}
\thanks{T.~Le is with the School of Engineering, Architecture and Aviation, Hampton University, Hampton, VA 23669, USA. Email: tan.le@hamptonu.edu.}
\thanks{V.~Le is with the Virginia Polytechnic Institute and State University, Blacksburg, VA 24061, USA. Email: vanl@vt.edu.}
\thanks{S.~Shetty is with the Virginia Modeling, Analysis and Simulation Center, Old Dominion University, Suffolk, VA 23435, USA. 
Emails: sshetty@odu.edu.}}
\maketitle

\begin{abstract}
\boldmath
Indoor robotic systems within Cyber–Physical Systems (CPS) are increasingly exposed to Denial of Service (DoS) attacks that compromise localization, control and telemetry integrity. 
We propose a privacy-aware malware detection framework for indoor robotic systems, which leverages hybrid quantum computing and deep neural networks to counter DoS threats in CPS, while preserving privacy information. 
By integrating quantum-enhanced feature encoding with dropout-optimized deep learning, our architecture achieves up to 95.2\% detection accuracy under privacy-constrained conditions. 
The system operates without handcrafted thresholds or persistent beacon data, enabling scalable deployment in adversarial environments. 
Benchmarking reveals robust generalization, interpretability and resilience against training instability through modular circuit design. 
This work advances trustworthy AI for secure, autonomous CPS operations.
\end{abstract}

\section{Introduction}
As robotic autonomy expands into privacy-sensitive indoor environments, the threat of real-time Denial of Service (DoS) attacks demands a new class of interpretable, quantum-enhanced AI defenses. The cybersecurity of \textit{Cyber–Physical Systems} (CPSs) has become increasingly critical across domains such as healthcare, defense and industrial automation \cite{kayan2022cybersecurity}. 
These systems tightly couple computation, control and physical processes, making them uniquely vulnerable to adversarial threats that traverse digital-physical boundaries and induce real-world harm. 
In particular, Indoor Positioning Systems (IPSs) used in autonomous robotics are susceptible to spoofing, jamming, and telemetry manipulation—conditions that degrade localization, disrupt control loops, and compromise mission integrity.

Traditional \textit{Intrusion Detection Systems} (IDSs) often fall short in autonomous platforms, as they overlook mobility, sensor fusion and real-time constraints. 
Recent work has explored onboard anomaly detection using telemetry and statistical change-point methods such as \textit{Cumulative Sum} (CUSUM)~\cite{bonczek2022detection} and physics-based models for sensor spoofing resilience~\cite{pu2022security}.
A core component of autonomous robotics is the \textit{Real-Time Location System} (RTLS) \cite{10538229}, which enables robots to localize, plan and execute tasks. 
As a result, many systems rely on external RTLSs, which are vulnerable to cyber-physical attacks.
Here, RTLSs can be implemented using technologies such as \textit{Global Positioning System} (GPS), \textit{Ultra-Wideband} (UWB), Wi-Fi and ultrasound. 
GPS spoofing has been widely studied, but the cybersecurity of IPSs remains underexplored. 
IPSs estimate position using signal properties such as time of flight, signal strength, angle of arrival and hop count~\cite{singh2024systematic}. 
These systems are susceptible to attacks including: 1) Forced multipath propagation; 2) Speedup and delay injection; 3) Replay and signal modification and 4) Jamming and Denial of Service (DoS)~\cite{singh2024systematic}.

Among IPS technologies, UWB-based systems are widely adopted in mobile robotics due to their high accuracy and low computational overhead \cite{zhao2021learning, flueratoru2021high, niculescu2022energy}. 
These systems estimate position by measuring the range between a mobile transceiver and multiple fixed beacons. 
However, beacon exposure makes them vulnerable to spoofing and jamming, which can corrupt localization estimates and compromise robotic navigation.
These vulnerabilities resemble those in \textit{Wireless Sensor Networks} (WSNs), where anomaly detection methods have been proposed to identify abrupt changes in communication patterns. 
For example, \cite{o2014anomaly} uses distributed CUSUM-based detection, while~\cite{6560057} introduces \textit{Verifiable Multilateration} (VM), a statistical method for position verification using known reference points. 
However, VM requires ground-truth beacon positions during operation, limiting its scalability.



To address these unique challenges, we propose the \textbf{Privacy-aware Framework of Robust Malware Detection on Indoor Robots} by using \textbf{Hybrid Quantum Computing and Deep Neural Networks}. 
This framework integrates deep learning with quantum-enhanced computing to detect malware attacks in indoor robotic systems. 
Our hybrid architecture leverages quantum feature encoding and variational quantum circuits (VQCs) to improve detection accuracy, interpretability and resilience against adversarial threats~\cite{beer2020training, hybrid_ai_uchicago, le2022artificial}.
Specifically, we focus on developing models for: 1) Intrusion detection and classification; 2) Event-triggered control and autonomous response; and 3) Privacy-preserving malware analysis.
Unlike traditional AI models, our hybrid quantum-classical system benefits from: 1) Exponential speedup in feature space exploration via quantum superposition; 2) Improved generalization in high-dimensional, noisy environments; and 3) Enhanced interpretability through quantum explainability frameworks.
Our numerical results show that hybrid quantum models can outperform classical baselines in malware detection tasks, achieving up to 95.238\% accuracy and F1 scores above 0.95 in adversarial settings. 
These results demonstrate the promise of quantum-enhanced AI in securing CPS environments, especially where real-time decision-making and privacy preservation are critical.
Our framework lays the foundation for secure, autonomous operation in indoor robotics, enabling proactive threat detection and adaptive control.
It contributes to the broader goal of trustworthy AI in CPS, combining modularity, reproducibility and ethical impact.

\section{Related Work}

\begin{table*}[htbp]
\centering
\caption{Comparison of This Study with Representative Literature}
\label{tab:comparison}
\begin{tabular}{|p{4.5cm}|p{5.3cm}|p{6.0cm}|}
\hline
\textbf{Feature} & \textbf{Prior Work} & \textbf{This Study} \\
\hline
\textbf{Target Domain} & Generic CPS or IoT malware detection~\cite{tifs_dl_malware,tcyber_dl_cps,ai_iomt_springer} & Robotic CPSs with real-time constraints and adversarial telemetry \\
\hline
\textbf{Model Type} & Classical DNNs (CNN, LSTM, RNN)~\cite{tifs_dl_malware,tcyber_dl_cps} & Hybrid deep quantum neural networks with variational circuits~\cite{kukliansky2024network,joshi2025quantumai} \\
\hline
\textbf{Quantum Integration} & Absent or limited to toy models~\cite{hybrid_ai_uchicago,quantum_ai_dive} & Full QNN pipeline with remote NISQ execution and telemetry encoding~\cite{vts2021nisq,broughton2021tensorflow} \\
\hline
\textbf{Explainability} & Limited interpretability; post-hoc analysis~\cite{tcyber_dl_cps} & Integrated explainability overlays via QuXAI~\cite{quxai_arxiv} \\
\hline
\textbf{Privacy Preservation} & Federated learning or encrypted features in generic IoT~\cite{tcyber_privacy_cps} & Privacy-aware telemetry ingestion tailored to robotic CPSs using lightweight data sanitization \cite{le2022artificial}\\
\hline
\textbf{Benchmarking and Reproducibility} & No standardized quantum benchmarking for CPS malware detection~\cite{mcclean2018barren} & Modular, reproducible QNN benchmarking pipeline with adversarial drift scenarios \\
\hline
\textbf{Deployment Context} & Onboard DNN inference or cloud-based classical machine learning (ML)~\cite{ieee_cps_ml_review} & Remote NISQ execution with hybrid control loop integration~\cite{preskill2018nisq} \\
\hline
\end{tabular}
\vspace{-0.3cm}
\end{table*}

Cybersecurity in \textit{Cyber–Physical Systems} (CPSs) has traditionally relied on rule-based intrusion detection and handcrafted anomaly monitoring~\cite{ieee_cps_intrusion}. While working effectively in static environments, these approaches often fail to generalize across dynamic, adversarial, and resource-constrained settings. Recent advances in machine learning have enabled data-driven detection of spoofing, jamming, and malware attacks, particularly in mobile and embedded robotics~\cite{ieee_cps_ml_review}. However, challenges remain in interpretability, robustness, and deployment feasibility.

\subsection{Malware Detection in Robotic CPSs}

Malware in robotic CPSs poses unique challenges due to tight coupling between sensing, actuation, and control. Unlike generic IT systems, robotic platforms operate under real-time constraints, limited compute budgets, and safety-critical feedback loops. Malware may exploit vulnerabilities in sensor fusion, control logic or wireless telemetry, leading to physical misbehavior rather than just data corruption~\cite{ieee_robotics_security}.
Traditional malware detection methods often assume static network topologies or abundant computational resources, which do not hold in mobile robotic platforms. Moreover, robotic CPSs frequently operate in partially observable environments, where telemetry is noisy, incomplete, or spoofed. These constraints demand lightweight, interpretable, and spoof-resilient detection mechanisms tailored to robotic workflows~\cite{qmlids2024}.

\subsection{Deep Neural Network–Based Malware Detection}

Deep learning models such as CNNs, RNNs, and LSTMs have shown promise in detecting malware signatures across CPS domains~\cite{tifs_dl_malware,tcyber_dl_cps}. In robotic contexts, these models are trained on telemetry traces, control sequences, or network traffic patterns to identify anomalous behaviors. However, several limitations persist:

\begin{itemize}
    \item \textbf{Data dependence:} DNNs often require large labeled datasets, which are difficult to obtain in robotic CPSs due to privacy, heterogeneity, and limited attack observability.
    \item \textbf{Overfitting risk:} High model capacity can lead to poor generalization in unseen environments, especially under adversarial drift.
    \item \textbf{Interpretability gap:} DNN decisions are typically opaque, making it difficult to trace malware attribution or validate safety-critical responses.
\end{itemize}
These limitations motivate the exploration of alternative architectures that offer better generalization, interpretability, and robustness under CPS constraints.

\subsection{Quantum Neural Networks for CPS Security}

Quantum Neural Networks (QNNs) have emerged as a promising alternative for malware detection in CPSs~\cite{kukliansky2024network,joshi2025quantumai}. By leveraging quantum feature encoding and variational circuits, QNNs can explore high-dimensional feature spaces more efficiently than classical models. This enables:
\begin{itemize}
    \item \textbf{Compact representations:} Quantum states encode complex correlations in fewer qubits, reducing memory overhead~\cite{schuld2020mlqc}.
    \item \textbf{Robustness to noise:} VQCs can be trained to tolerate decoherence and adversarial perturbations~\cite{moll2025qvc}.
    \item \textbf{Hybrid optimization:} Classical optimizers can tune quantum parameters via gradient-based feedback, enabling scalable training~\cite{cerezo2021variational}.
\end{itemize}
These advantages are particularly relevant in robotic CPSs, where telemetry is sparse, noisy, and adversarially vulnerable. QNNs offer a path toward lightweight, interpretable, and hardware-compatible malware detection. Recent implementations on IonQ’s Aria-1 quantum computer have demonstrated competitive performance in intrusion detection tasks, achieving an F1 score of 0.86 on benchmark datasets~\cite{kukliansky2024network}.

\subsection{Deployment Context: NISQ Hardware}

The proposed QNN framework targets deployment on \textit{Noisy Intermediate-Scale Quantum} (NISQ) devices, i.e. quantum processors with tens to hundreds of qubits, limited coherence times, and restricted gate fidelity~\cite{preskill2018nisq}. 
These devices are accessible via cloud-based platforms such as IBM Quantum, IonQ, and Rigetti, which offer remote execution of variational circuits.
In our context, the QNN is trained classically and executed on a remote NISQ backend, with telemetry features encoded into quantum states via amplitude or angle encoding. 
This hybrid setup allows robotic CPSs to offload malware detection to quantum co-processors, while maintaining real-time control locally. 
Although NISQ devices are not yet deployable onboard mobile robots, cloud-accessible quantum inference offers a viable path for near-term CPS security augmentation~\cite{vts2021nisq}.

\subsection{Explainability and Privacy Preservation}

Interpretability remains a critical challenge in AI-driven cybersecurity. Frameworks such as QuXAI~\cite{ieee_quxai} provide explainability for hybrid quantum models, enabling transparent decision-making in safety-critical applications. Concurrently, privacy-preserving techniques such as federated learning and encrypted feature extraction are being explored to protect sensitive robotic telemetry~\cite{tcyber_privacy_cps}, especially in distributed CPS environments.

\subsection{Our Contribution} 
To clarify the novelty and scope of our proposed framework, the main contributions of this study are itemized below and contrasted with representative literature in Table~\ref{tab:comparison}:
\begin{itemize}
    \item We target \textbf{robotic CPSs} operating under real-time constraints and adversarial telemetry, rather than generic IoT or industrial CPS settings.
    \item We introduce a \textbf{hybrid deep quantum neural network} architecture tailored for malware detection, leveraging VQCs and telemetry encoding.
    \item We implement a \textbf{full QNN pipeline} with remote execution on NISQ hardware, integrating telemetry ingestion, quantum feature encoding, and inference.
    \item We embed \textbf{explainability overlays} using QuXAI to enhance interpretability in safety-critical decision-making.
    \item We incorporate \textbf{privacy-aware telemetry handling} by using data sanitization \cite{le2022artificial}, addressing CPS-specific constraints in federated and encrypted data flows.
    \item We establish a \textbf{modular benchmarking framework} for reproducible evaluation under adversarial drift and spoofing scenarios.
\end{itemize}
These contributions collectively advance the state of the art in quantum-enhanced CPS security, as summarized in Table~\ref{tab:comparison}.
In short, we build on these foundations by integrating quantum-enhanced malware detection with privacy-aware control and explainability overlays. 
Our framework is tailored to robotic CPSs, emphasizing reproducibility, modularity, and ethical impact, while remaining compatible with real-world deployment constraints.

\section{Problem Formulation and Quantum Computing Deep Neural Network Solutions}

\subsection{System Model of Robot Attack Detection}

We build a rigorous RTLS anomaly benchmark, especially for indoor robotics.
In particular, we derive both the evaluation framework and the feature transformations that allow fair, reproducible and privacy-safe analysis.
The scenario considered includes 1) developing the computing algorithm to detect the possible attacks, and 2) preserving privacy when sharing the dataset to avoid malicious actors to keep track on exposing user footprint.
It implies that we build a preprocessing module to anonymize the sensitive features before training, or simulate the impact of masking certain features on attack F1 Score. 
The data collection includes 10 features, i.e.
$ \mathbf{x} = [x_1, x_2, \dots, x_{10}] \in \mathbb{R}^{10}$.
These ten features and their privacy sensitivity as well as their description of possible risk and handling can be briefly summarized as follows:
\begin{itemize}
\item $x_1$ is the RSSI Mean, which reflects signal energy, but not identity. So, the privacy sensitivity is low.
\item $x_2$ is the RSSI Std Dev, which is informative for detection scheme but non-identifying for privacy leak. Hence, its privacy risk is low.  
\item $x_3$ is the Timestamp Jitter. Attackers can infer behavioral timing and/or recommend time windowing. Hence, its privacy risk is moderate.
\item $x_4$ is the Distance Estimate, which can reveal proximity patterns. Hence, it is considered a obfuscating absolute value. Its privacy sensitivity would be Moderate–High.
\item $x_5$ is the Positional Jitter. Malicious actors can use it to reconstruct movement paths, obfuscate and/or aggregate. So, its privacy sensitivity would be High.
\item $x_6$ is the Beacon ID Count/Entropy, which is directly linked to identifiable transmitters. Meaning that its privacy sensitivity is High.
\item $x_7$ is the Packet Drop Rate, which is only a behavioral signal. So, it gives low re-identification risk.
\item $x_8$ is the Anchor Signal Variance. This is an abstract signal pattern and hence is not user-tied. So, its privacy risk is low.
\item $x_9$ is the Estimated Velocity, which could imply user/robot behavior. Its privacy sensitivity is Moderate and we can discretize or anonymize it when we need to handle this possible risk.
\item $x_{10}$ is the Velocity Residual vs Odometry, which is linked to movement profiling.  Its privacy sensitivity is Moderate and we need to mask odometry source if needed.
\end{itemize}
The label for the dataset denotes by $y$ ($y \in {0,1,2}$), which represents normal operation or DOS attack or spoofing attack. 
Our goal is to build a model $f_\theta(\mathbf{x}) \rightarrow y$ that detects anomalies, whilst respecting privacy constraints.

We now calculate these parameters as follows.
In RTLS systems of indoor robots, when a beacon broadcasts a Bluetooth signal, the receiving device (e.g. the robot or sensor) measures the signal strength (RSSI).
Here, the RSSI measures the power level of a received radio signal. 
The relationship between signal strength and distance can be used to estimate how far away a beacon is from the receiver.
Hence, RSSI values is useful for estimating the proximity, signal reliability, and beacon identity of the users/robots/sensors.

While raw RSSI is device-calibrated, the general form is
$ \text{RSSI} = P_{rx} = P_{tx} - PL(d) $, where $P_{tx}$ is the transmit power, $PL(d)$ is the path loss at distance $d$.
We can also model RSSI with $ RSSI(d) = -10n \cdot \log_{10}(d) + C $, where $n$ is the path loss exponent (2–4 for indoor) and $C$ is the calibration constant (device-specific).
We can derive multiple metrics of derivatives from RSSI, called RSSI Stats, which are used to capture attack patterns, given as
\begin{itemize}
\item Mean RSSI: Average signal strength. The Spoofing attacker may elevate or suppress it.
\item RSSI Std Dev: It measure fluctuation in signal. The DoS attacker or jammer (jamming attack) causes instability.
\item RSSI Entropy or Beacon Signal Entropy: It measure the diversity in RSSI values per window. The Spoofer may inject consistent signals.
\item RSSI Drop Rate: It counts of missing packets or low-RSSI packets. DoS attacks can manifest as a sudden and unexplained drop in RSSI, an increase in missing packets, or a higher proportion of packets received with very low RSSI values. This may be due to the attacker jamming the channel or creating significant interference, hindering the delivery of legitimate network traffic. It is called as the DoS attack signature.
\end{itemize}
In short, these related features derived from RSSI give the benchmark sensitivity to spoofing and DoS behaviors.
Hence, they would be fed directly into our proposed classifier as attack indicators.

As we observed that RSSI alone is moderately safe in the perspective of privacy sensitivity because it does not directly identify users or devices.
However, when it is paired with positional data, it can triangulate locations. 
Also, persistent patterns in RSSI per beacon may allow re-identification.
Moreover, its privacy risk grows when it is combined with Beacon IDs for device-level tracking, with precise timestamps for tracking behavioral patterns, and with location estimates for path reconstruction via triangulation.
So, we must consider the following essential strategies in privacy-preserving workflows.
1) Using RSSI categories or quantized bins instead of exact dBm; 2) Aggregating RSSI over zones rather than per beacon; 3) Masking device-level RSSI traces if beacon IDs are sensitive. 

Let us derive these essential features as follows:
\begin{itemize}
\item RSSI Stats: $\mu_{RSSI} = \frac{1}{N}\sum_{i=1}^N RSSI_i$, $\sigma_{RSSI} = \sqrt{\frac{1}{N}\sum(RSSI_i - \mu)^2}$.
\item Timestamp Jitter: $\Delta t_i = t_{i} - t_{i-1}$; variance computed over window.
\item Distance Estimate: TOA: $d = c \cdot \Delta t$; TDOA: $d_{A} - d_{B}$.
\item Positional Jitter: $\sigma_{loc} = \sqrt{\text{Var}(x) + \text{Var}(y)}$.
\item Beacon Entropy: $H = -\sum p_i \log_2 p_i$, where $p_i$ is frequency of beacon ID $i$.
\item Velocity: Euclidean change over time; residual compared to robot's IMU or wheel encoders.
\end{itemize}

\subsection{Problem Statement}
\subsubsection{Detection of Malicious Actor without Considering Privacy Preservation}
We aim to develop a supervised learning model for detecting cyber-physical attacks on indoor robotic systems using RTLS telemetry. 
The model ingests both quantitative and categorical features and outputs a discrete label: 1) \textbf{WA}-No Detection of attack detected without Considering Privacy Preservation; 2) \textbf{A}-Attack detected.
Given the categorical nature of the output, classification algorithms are preferred over regression-based predictors. 
Our pipeline supports both raw and privacy-transformed feature ingestion, enabling ethical deployment in multi-tenant edge environments.
Especially, we will compare our proposal with the regular NN, DNN, and CNN \cite{Zahin19, Zahin20, Wang20,le2025dpfaga}.
The class-wise performance metrics used in evaluation include accuracy, precision, recall and different F$_1$ score, with emphasis on validation accuracy to assess generalization.

\subsubsection{Privacy-Preserving Feature Transformations}

To support ethical deployment in shared edge environments, we apply privacy-aware transformations to sensitive features. These transformations obscure user-specific telemetry while retaining attack-relevant signals:
\begin{itemize}
    \item \textbf{Zone-Level Encoding ($x_4$, $x_5$)}: Replace continuous coordinates with discrete zones (e.g., room, sector).
    \item \textbf{Beacon ID Hashing ($x_6$)}: Rotate anonymized beacon identifiers periodically.
    \item \textbf{Velocity Discretization ($x_9$, $x_{10}$)}: Map velocity to movement categories: ``stationary'', ``slow'', ``fast''.
    \item \textbf{Timestamp Bucketization ($x_3$)}: Aggregate timestamps into coarse intervals (e.g., 1-minute blocks).
\end{itemize}

For instance, we define:
\begin{itemize}
    \item Privacy-sensitive subset: $\mathcal{S}_p \subseteq \{x_3, x_4, x_5, x_6, x_9, x_{10}\}$
    \item Attack-relevant subset: $\mathcal{S}_a \subseteq \{x_1, x_2, x_7, x_8\}$
\end{itemize}
Let $T_p(\mathbf{x})$ denote the privacy-aware transformation. 
We benchmark detection fidelity as:
\[
\text{Attack F1}(f_\theta(T_p(\mathbf{x}))) \quad \text{vs} \quad \text{Attack F1}(f_\theta(\mathbf{x}))
\]
This quantifies the trade-off between privacy preservation and detection accuracy.

\subsection{Hybrid Quantum Deep Neural Network Architecture}

\begin{figure}[!t]
\centering
\includegraphics[width=0.85\linewidth]{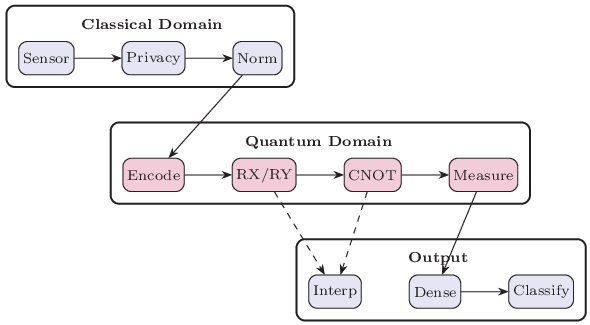} 
\caption{\scriptsize Hybrid quantum-classical architecture with staggered layout.}
\vspace{-0.3cm}
\label{fig:staggered_qnn_eps}
\end{figure}




To overcome limitations of classical models in high-dimensional, noisy and adversarial environments, we propose a hybrid deep quantum neural network (DQNN) architecture. This system integrates:
\begin{itemize}
    \item \textbf{Quantum Feature Encoding}: Maps classical features into quantum states using amplitude or angle encoding.
    \item \textbf{VQCs}: Learn nonlinear decision boundaries via tunable quantum gates.
    \item \textbf{Deep Learning Layers}: Handle preprocessing, feature selection, and post-quantum classification.
\end{itemize}
Our architecture of quantum-enhanced malware detection and explainability  supports:
\begin{itemize}
    \item \textbf{Exponential feature space exploration} via quantum superposition
    \item \textbf{Improved generalization} in adversarial and noisy conditions
    \item \textbf{Enhanced interpretability} through quantum circuit visualization and attribution
\end{itemize}
Also, our proposed framework is modular, reproducible and privacy-aware and can support:
\begin{itemize}
    \item \textbf{Attack detection and classification}
    \item \textbf{Event-triggered control and autonomous response}
    \item \textbf{Privacy-preserving telemetry ingestion}
    \item \textbf{Quantum-enhanced interpretability}
\end{itemize}
This architecture lays the foundation for secure, autonomous operation in indoor robotics and CPS environments. 
It contributes to the broader goal of trustworthy AI—balancing performance, interpretability and ethical safeguards.

\section{Hybrid Deep Quantum Neural Networks and Deep Neural Networks for Privacy-Aware Attack Detection}
\label{DQCNN}

\subsection{Quantum Machine Learning}
\label{subsec:QML}

Machine learning (ML) involves constructing algorithms that learn patterns from data to make predictions on unseen inputs. 
While early ML research emphasized theoretical guarantees \cite{murphy2012machine}, recent advances have favored heuristic methods like deep learning \cite{goodfellow2016deep}, which learn representations via parameterized networks optimized through loss functions.
In parallel, quantum computing has gained momentum due to its ability to simulate phenomena such as superposition and entanglement \cite{preskill2018simulating}. 
Quantum computers promise speedups in domains including chemistry, cryptography, and optimization \cite{farhi2014quantum}. 
Quantum Machine Learning (QML) explores how quantum systems can accelerate ML tasks. 
First-generation QML algorithms leverage quantum linear algebra to speed up classical tasks such as principal component analysis, support vector machines, clustering and recommendation systems. 
However, embedding classical data into quantum states remains a scalability challenge and quantum speedups are often constrained by data structure \cite{huang2021power}.
With the rise of Noisy Intermediate-Scale Quantum (NISQ) devices \cite{preskill2018quantum}, a second generation of QML has emerged. 
These models use parameterized quantum circuits (PQCs), also known as Quantum Neural Networks (QNNs), trained via gradient-based or heuristic optimization \cite{chen2021universal}. 
This mirrors the evolution of classical ML toward deep learning, driven by increased computational power.
QML now focuses on designing quantum-native models, training strategies, and inference schemes that exploit quantum properties for learning tasks.
Specially, each qubit in QML undergoes the tasks of data encoding, rotation gates, entanglement and measurement.


\subsection{Efficient Learning for Deep Quantum Computing Neural Networks}
\begin{figure}[htbp]
\vspace{-0.3cm}
    \centering
    \includegraphics[width=0.85\linewidth]{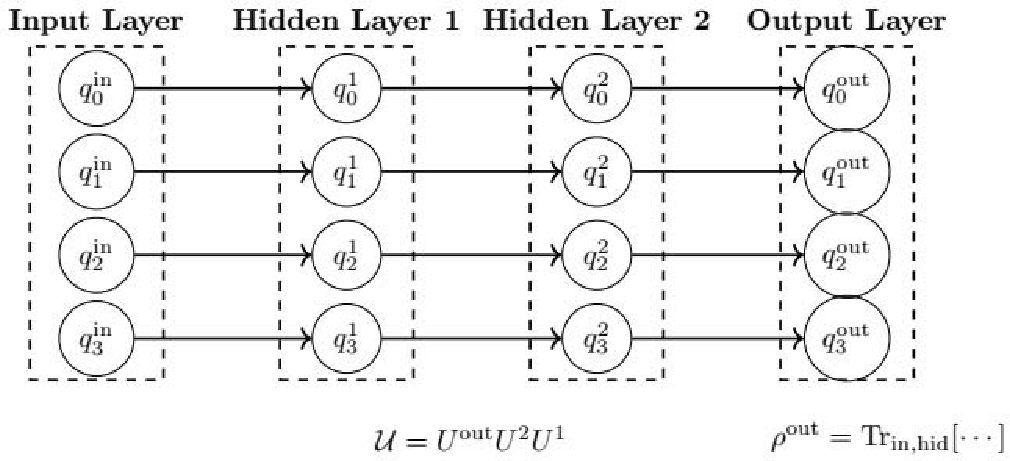}
    \caption{Quantum neural network architecture with input, hidden, and output layers. Each arrow represents a quantum perceptron unitary $U_j^l$ acting between layers.}
    \vspace{-0.3cm}
    \label{fig:qnn_architecture_fig}
\end{figure}
\begin{figure}[htbp]
\vspace{-0.3cm}
    \centering
    \includegraphics[width=0.85\linewidth]{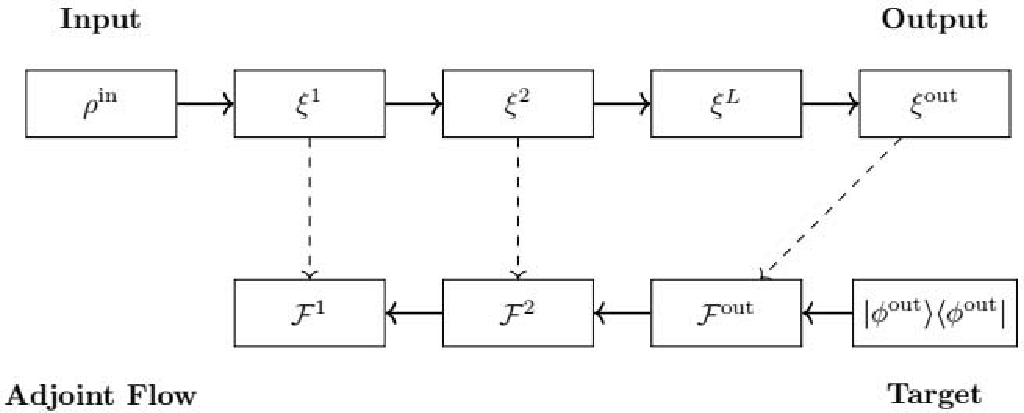}
    \caption{Quantum backpropagation flow. Forward propagation of $\rho^{\text{in}}$ through CP maps $\xi^l$, and backward propagation of the target state via adjoint channels $\mathcal{F}^l$. Dashed arrows indicate layer-wise gradient computation.}
    \vspace{-0.3cm}
    \label{fig:qnn_backprop}
\end{figure}
\subsubsection{Quantum Perceptrons and Network Architecture}
There are critical challenges, when designing QML algorithms for quantum data, including: 1) identifying a quantum generalization of the perceptron, 2) constructing deep neural network architectures, 3) specifying loss functions, and 4) developing optimization strategies. 
We address these by proposing a natural quantum perceptron which is integrated into a QNN to enable universal quantum computation.

We develop our QNN architecture shown in Figs. \ref{fig:qnn_architecture_fig} and \ref{fig:qnn_backprop}, which is the modification from the quantum feedforward neural networks \cite{beer2020training} to the new quantum backpropagation neural networks flow. 
In particular, our model supports a quantum analogue of classical backpropagation by leveraging completely positive (CP) layer transition maps. 
We apply this framework to the task of learning an unknown unitary transformation in both ideal and noisy conditions.
Classical simulations suggest that our method is feasible for NISQ devices.

Several quantum perceptron models have been proposed, including circuit-based qubit setups and continuous-variable systems \cite{schuld2020circuit,torrontegui2019unitary}. 
Our model generalizes these in the same way as quantum feedforward neural networks \cite{beer2020training}, i.e. by defining a quantum perceptron as an arbitrary unitary acting on $m$ input qubits and $n$ output qubits. 
The input is initialized in a mixed state $\rho^{\text{in}}$, while the output in a fiducial product state $|0\cdots0\rangle_{\text{out}}$. 
The perceptron unitary depends on $(2^{m+n})^2 - 1$ parameters, including weights and biases.
In the following, we briefly present the QNN, interested readers can find detailed descriptions and derivations in \cite{beer2020training}.

\subsubsection{Quantum Neural Network Construction}
We begin by introducing a basic class of quantum perceptrons known as controlled-unitary perceptrons \cite{torrontegui2019unitary}. These gates apply a unitary transformation conditioned on the classical basis state of a control register:
\beqn
U = \sum_{\alpha} |\alpha\rangle \langle\alpha| \otimes U(\alpha),
\eeqn
where $|\alpha\rangle$ spans the input basis and $U(\alpha)$ are parameterized unitaries. 
When applied to a quantum state, this structure yields a classical-quantum (CQ) channel, $\rho^{\text{out}} = \sum_{\alpha} \langle \alpha| \rho^{\text{in}} |\alpha\rangle \, U(\alpha) |0\rangle \langle0| U(\alpha)^\dagger.$
Such channels collapse quantum coherence in the control register and have zero quantum channel capacity. 
While conceptually simple, controlled-unitary perceptrons cannot support general quantum computation and are unsuitable for tasks requiring entanglement propagation, quantum memory, or adversarial robustness.

To overcome these limitations, we adopt the QNN, i.e. a layered quantum architecture that generalizes perceptron behavior and enables full quantum expressivity \cite{beer2020training}. 
The QNN consists of $L$ layers of perceptrons acting on an initial quantum state $\rho^{\text{in}}$, along with ancillary qubits initialized in the state $|0\cdots0\rangle_{\text{hid,out}}$. 
The full system undergoes unitary evolution via a circuit $\mathcal{U}$, which entangles the input, hidden, and output registers. 
To extract the final output state, we apply a partial trace over the input and hidden subsystems by effectively discarding them and retaining only the reduced density matrix of the output, given as
\beqn
\rho^{\text{out}} = \text{Tr}_{\text{in,hid}} \left( \mathcal{U} \left( \rho^{\text{in}} \otimes |0\cdots0\rangle_{\text{hid,out}}\langle0\cdots0| \right) \mathcal{U}^\dagger \right).
\eeqn
Here, $\text{Tr}_{\text{in,hid}}$ denotes the partial trace, a mathematical operation that removes degrees of freedom associated with the input and hidden registers. This is essential in quantum modeling, as it allows us to focus on the observable output while marginalizing over internal states that are not measured. 
Unlike a full trace, which yields a scalar, the partial trace returns a valid quantum state over the remaining subsystem in this case, the output register.

Moreover, $\mathcal{U} = U^{\text{out}} U^L \cdots U^1$ is the full QNN circuit, and each $U^l$ is a product of perceptrons acting between layers l-1 and l. 
Unlike controlled-unitary gates, these perceptrons operate on entangled registers and ancilla, preserving coherence and enabling arbitrary quantum channel construction. 
Since the unitaries generally do not commute, the order of operations is significant and contributes to the network's expressivity.
In practice, we construct QNNs using noncommuting perceptrons acting on qubit registers. 
These gates are well-suited to current quantum hardware platforms and offer several key advantages. 
Their noncommutativity enables rich entanglement dynamics across layers, which is essential for learning complex quantum transformations. 
When combined with ancilla initialization and partial trace operations, these perceptrons can implement arbitrary CP maps. 
This makes the QNN architecture highly expressive—capable of efficiently representing both unitary and non-unitary quantum processes. 
Its structure supports gradient-based optimization, enables layer-wise interpretability and remains resilient under noise, decoherence and adversarial environments.
These notable benefits make QNN well-suited for real-world quantum learning and control tasks.

The QNN output can also be expressed as a composition of completely positive maps, i.e. $\rho^{\text{out}} = \xi^{\text{out}} \circ \xi^L \circ \cdots \circ \xi^1 (\rho^{\text{in}}),$ where each $\xi^l$ is defined by $\xi^l(X^{l-1}) = \text{Tr}_{l-1} \left( \prod_{j=m_l}^1 U_j^l \left( X^{l-1} \otimes |0\cdots0\rangle\langle0\cdots0| \right) \prod_{j=1}^{m_l} {U_j^l}^\dagger \right),$ with $U_j^l$ denoting the $j$th perceptron in layer $l$, and $m_l$ the number of perceptrons in that layer. 
This feed-forward structure enables a quantum version of backpropagation and supports interpretability overlays for benchmarking and visualization.

\subsubsection{Learning Unknown Unitaries}

We consider training data consisting of pairs $(|\phi^{\text{in}}_x\rangle, |\phi^{\text{out}}_x\rangle)$ for $x = 1,\dots,N$, where $|\phi^{\text{out}}_x\rangle = V |\phi^{\text{in}}_x\rangle$ for some unknown unitary $V$. 
This models scenarios, where an uncharacterized device performs a quantum operation on arbitrary inputs.
To evaluate QNN performance, we use fidelity as the cost function, i.e.
\beqn
\mathcal{C} = \frac{1}{N} \sum_{x=1}^N \langle \phi^{\text{out}}_x | \rho^{\text{out}}_x | \phi^{\text{out}}_x \rangle,
\eeqn
where $\rho^{\text{out}}_x$ is the QNN output for input $|\phi^{\text{in}}_x\rangle$. 
The fidelity ranges from 0 (worst) to 1 (best). 
For mixed output states, the cost function generalizes accordingly.

\subsubsection{Quantum Backpropagation and Optimization}

In the training step, we update each perceptron unitary via
\beqn
U \mapsto e^{i\varepsilon \mathcal{K}} U,
\eeqn
where $\mathcal{K}$ encodes the update direction and $\varepsilon$ is the step size. 
The change in cost is
\beqn
\Delta \mathcal{C} = \frac{\varepsilon}{N} \sum_{n=1}^N \sum_{l=1}^L \text{Tr} \left[ \chi_n^l \Delta \xi^l(\rho_n^{l-1}) \right],
\label{eq:qnn_cost_update}
\eeqn
where $\rho_n^{l-1}$ is the state from previous layers, $\chi_n^l$ is the backpropagated adjoint state from the output, $\mathcal{F}(X) = \sum_\pi A_\pi^\dagger X A_\pi$ is the adjoint channel of the CP map $E(X) = \sum_\pi A_\pi X A_\pi^\dagger$.

To compute $\mathcal{K}_j^l$ for a specific perceptron, we only require:
\begin{itemize}
    \item The output state of the previous layer, $\rho^{l-1}$,
    \item The adjoint-propagated state $\chi_n^l$ from the desired output.
\end{itemize}
This layer-wise update avoids applying the full QNN unitary across all qubits, reducing memory requirements and enabling scalable training of deep QNNs. 
Matrix sizes scale only with network width, but not network depth.

\subsection{Privacy-Preserving Strategies}
\begin{figure}[!t]
\vspace{-0.3cm}
\centering
\includegraphics[width=0.65\linewidth]{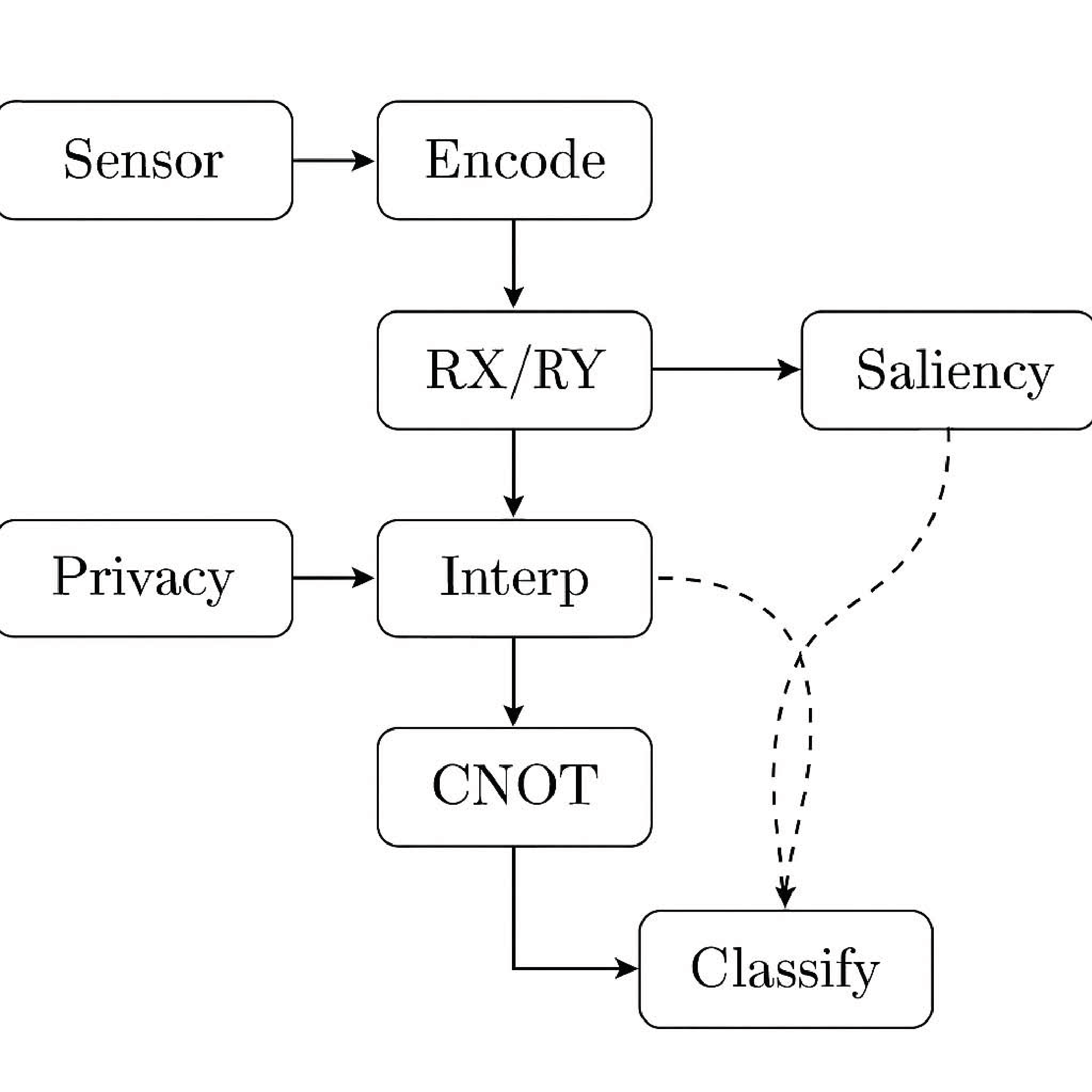} 
\caption{Spoof-Aware Annotation: Spoof-aware annotation showing adversarial injection points in the
system.}
\vspace{-0.3cm}
\label{fig:staggered_qnn_eps}
\end{figure}
Our considered scenario is that detection scheme would be performed by the pool of edge computing nodes.
Note that our indoor robot platform is integrated in the multi-tenant edge computing environments, in which different applications and/or their users share the same infrastructure of data storage, data processing, control operations, etc.
Therefore, there is a risk that malicious actors may be in this computing pool and can attempt to access or extract private data belonging to other users/robots. 
This leads to a critical security problem, called the risk of data ex-filtration. 
For instance, the exposed sensitive data can lead to data breaches and the attackers can bypass external-facing defenses and exploit vulnerabilities to extract data.
Furthermore, the malicious actors are able to compromise the privacy of individuals and organizations within the common infrastructure. 
To address these critical challenges, we need to perform data sanitization before sharing these data for training and testing (see Fig. \ref{fig:staggered_qnn_eps}). 
Similar to our previous data sanitization \cite{le2022artificial}, we briefly describe our four strategies of privacy preservation for features of robot data as follows:
\begin{itemize}
\item \textbf{Replace Precise Position Features ($x_4$ and $x_5$) with Zone-Level Encodings}: Instead of feeding continuous distance or jitter values, we encode whether the signal crosses zones or deviates from expected anchors.
\item \textbf{Mask or Hash Beacon IDs (Feature $x_6$) with Temporal Rotation}: We use anonymized IDs that rotate periodically to prevent persistent tracking.
\item \textbf{Discretize Velocity Estimates (Features $x_9$ and $x_{10}$) into movement categories}: For example, we formulate the new domain with three bases of movement of ``stationary'', ``slow'', and ``fast'' as well as multiple levels for each moving base.
Then, we will project the current velocity estimates to the established domain.
This procedure retains anomaly signal but obscures fine behavior.
\item \textbf{Time Bucketization for Timestamps (Feature $x_3$)}: For example, in the DoS attack scenario, we group timestamps into the coarse intervals (e.g. 1-minute blocks) to hide usage patterns, while keeping DoS sensitivity.
\end{itemize}
These proposed models could ingest these transformed features directly, supporting attack detection without exposing user footprint, called a privacy-preserving RTLS defense, that balances accuracy and ethics.

So now, we perform the feature categorization, where the feature set is split as
\begin{itemize}
\item Privacy-sensitive feature subset: $(\mathcal{S}_p \subseteq [x_3, x_4, x_5, x_6, x_9, x_{10}])$
\item Attack-relevant feature subset: $(\mathcal{S}_a \subseteq [x_1, x_2, x_7, x_8])$
\end{itemize}
We define the transformation functions as
\begin{itemize}
\item Privacy-aware transform: $T_p(\mathbf{x}) \rightarrow \tilde{\mathbf{x}}$
\item Benchmark fidelity is then measured as: $ \text{Attack F1}(f_\theta(T_p(\mathbf{x}))) \quad \text{vs} \quad \text{Attack F1}(f_\theta(\mathbf{x})) $
\end{itemize}
In practice, we can implement the Privacy-Preserving Strategies as follows: 1) Replace exact positions (features 4–5) with zone-level encodings: room, sector, grid; 2) Replace Beacon IDs with hashed IDs that rotate periodically; 3) Bucket velocity values into categories: ``static'', ``moving'', ``fast''; 4) Aggregate timestamps into coarse intervals (e.g. minute-level).

\subsection{Hybrid Deep Quantum Computing Neural Networks and Deep Neural Network Architecture for Robust Malware Detection and Privacy Preservation}

\subsubsection{NISQ Algorithms for Hybrid Quantum Neural Network Malware Detection}
\vspace{-0.3cm}
\begin{figure}[htbp]
    \centering
    \includegraphics[width=0.95\linewidth]{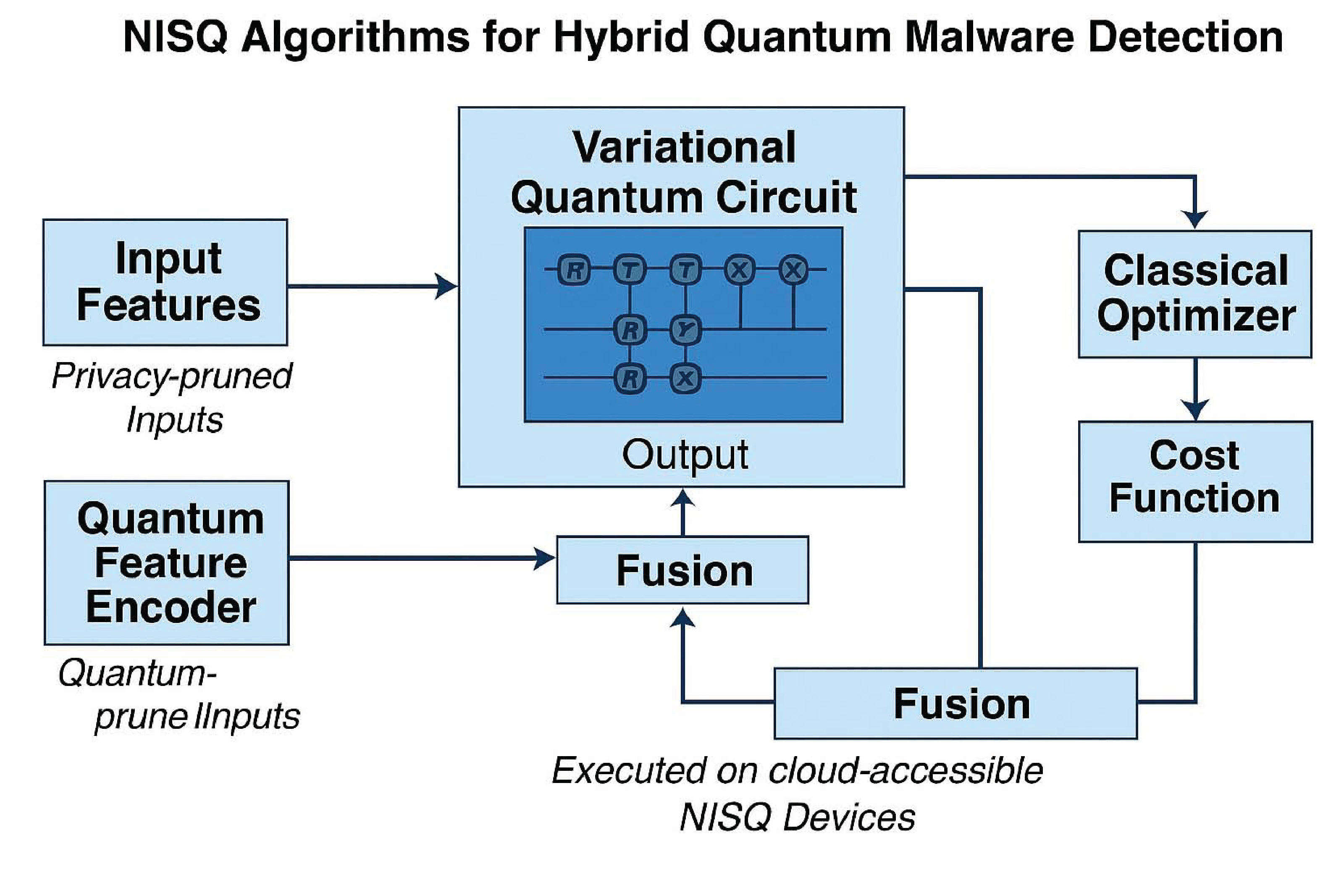}
    \caption{NISQ-compatible hybrid quantum malware detection pipeline. Input features are encoded into quantum states, processed by a VQC, and optimized via classical feedback. Outputs are fused with classical logic for robust and interpretable detection.}
    \vspace{-0.3cm}
    \label{fig:nisq_pipeline}
\end{figure}
NISQ algorithms are designed to operate within the constraints of near-term quantum hardware—devices with limited qubit counts, short coherence times and imperfect gate fidelity~\cite{preskill2018nisq}. Unlike fault-tolerant quantum algorithms, NISQ methods rely on shallow circuits and hybrid classical-quantum optimization to extract meaningful results despite noise.

In our VQC framework, we employ variational quantum algorithms such as the QNN and Quantum Approximate Optimization Algorithm to model adversarial uncertainty and detect malware signatures \cite{bharti2022noisy, preskill2018nisq}. 
Figure~\ref{fig:nisq_pipeline} illustrates the hybrid quantum-classical malware detection pipeline tailored for NISQ devices. 
The input layer ingests classical telemetry features, including privacy-pruned data after using our data sanitization \cite{le2022artificial}. 
These features are encoded into quantum states via amplitude and/or angle encoding in the Quantum Feature Encoder, preceded by normalization or masking. 
The encoded states are processed by a shallow-depth VQC composed of parameterized gates (e.g. $R_x$, $R_y$, $CZ$) and entanglement blocks arranged in linear or ring topologies. 
Measurement yields expectation values or bitstring samples, which are evaluated by the cost function,  i.e the expectation value of the Hamiltonian \cite{bharti2022noisy}. 
Our proposed optimization mechanisms used in \cite{6190772, 5986751, Tan2024} updates quantum parameters via gradient estimation techniques such as parameter-shift and finite-difference methods. 
The output is then fused with our DNN \cite{le2025dpfaga, Zahin19, Zahin20, Wang20} using a confidence-weighted fusion layer to enhance robustness and interpretability. The entire pipeline can be executed on cloud-accessible NISQ hardware (e.g. IBM Quantum, IonQ Aria-1), with support for circuit repetition and coherence-aware design to accommodate limited qubit depth.
These algorithms are particularly well-suited to NISQ platforms due to their following benefits, i.e. 
\begin{itemize}
    \item \textbf{Shallow circuit depth:} Reduces decoherence impact and enables execution on current superconducting and trapped-ion devices.
    \item \textbf{Parameterized unitaries:} Allow flexible encoding of malware features and adversarial perturbations.
    \item \textbf{Hybrid optimization loop:} Optimizers tune quantum parameters via cost function feedback, enabling scalable training~\cite{cerezo2021variational}.
\end{itemize}

We specifically design our QNN layers to minimize qubit overhead by using entanglement-efficient perceptrons and partial trace operations. This ensures compatibility with cloud-accessible NISQ backends such as IBM Quantum and IonQ Aria-1~\cite{vts2021nisq}. Furthermore, our architecture supports gradient estimation via parameter-shift rules and finite-difference methods, which are feasible on NISQ hardware with rapid circuit repetition.
By leveraging NISQ algorithms, our hybrid model achieves robust malware detection under realistic hardware constraints, while preserving interpretability and privacy. This positions our framework as a practical and forward-compatible solution for quantum-enhanced CPS security.

\subsubsection{Integration of Hybrid Deep Quantum Computing Neural Networks and Deep Neural Network Architecture}

\begin{figure}[htbp]
\vspace{-0.3cm}
    \centering
    \includegraphics[width=1\linewidth]{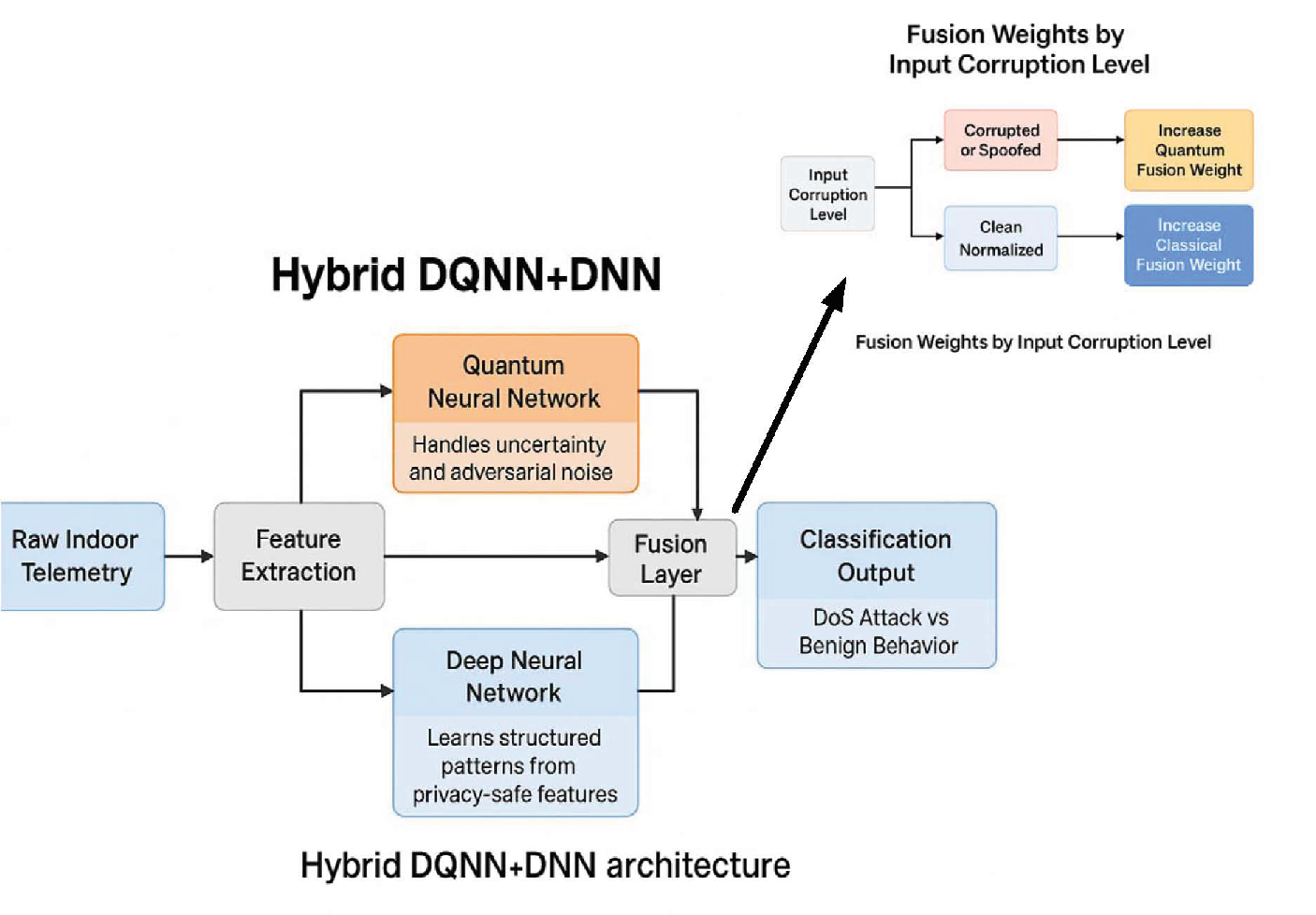}
    \caption{Hybrid quantum computing and deep neural network architecture for robust malware detection and privacy preservation.}
    \vspace{-0.3cm}
    \label{DQNNarchitecture}
\end{figure}

\begin{figure}[htbp]
    \centering
    \includegraphics[width=0.85\linewidth]{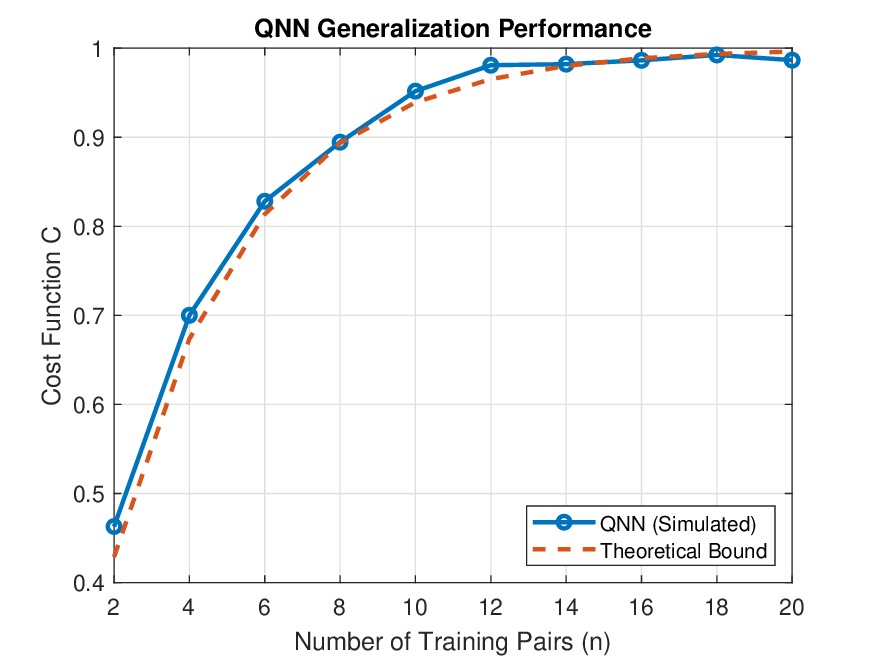}
    \caption{Generalization performance of QNN trained to learn a random unitary $V$. The cost function $C$ improves with the number of training pairs $n$, closely matching the theoretical bound.}
    \vspace{-0.3cm}
    \label{fig:qnn_generalization}
\end{figure}

\begin{figure}[htbp]
    \centering
    \includegraphics[width=0.85\linewidth]{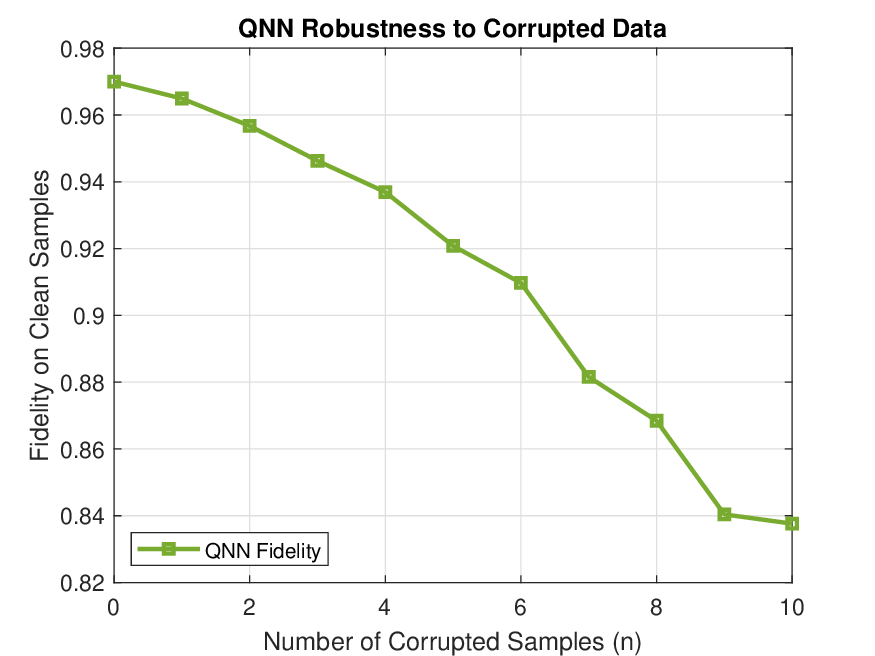}
    \caption{Robustness of QNN to corrupted training data. Fidelity on clean samples remains high even as the number of corrupted inputs increases.}
    \vspace{-0.3cm}
    \label{fig:qnn_robustness}
\end{figure}
We present the integration of Deep Quantum Neural Networks (DQNN) and Deep Neural Networks (DNN) for robust malware detection and privacy preservation in Fig.~\ref{DQNNarchitecture}. The architecture begins with telemetry inputs, where the system potentially privacy-pruned—and processes them through two parallel branches. The DQNN branch models uncertainty, entanglement and adversarial noise using shallow quantum circuits, making it resilient to spoofed or degraded signals. 
The DNN branch, implemented using CNN-LSTM layers, learns structured patterns from clean data and provides interpretable decision logic. 
These branches are unified via a trainable fusion layer that adaptively balances robustness and interpretability based on input quality. 
The fusion mechanism supports confidence-weighted blending, enabling dynamic trust assignment across modalities. 
This hybrid DQNN+DNN design allows cross-validation of signals, preserves privacy even with anchor features removed and maintains high attack detection performance. The architecture is optimized for NISQ-era constraints, using low qubit depth and efficient gradient estimation to ensure compatibility with cloud-accessible quantum platforms.
This architecture include three main components of DQNN, DNN and Fusion Layer, i.e. 
\begin{itemize}
    \item DQNN: this models uncertainty, entanglement and adversarial noise. Here, the quantum logic helps detect subtle anomalies that classical models might miss, especially in privacy-pruned inputs. As a result, this component 1) is robust to spoofed or degraded signals, 2) capture non-classical correlations in sensor data and 3) is useful when anchor data is sparse or partially obfuscated.
    \item DNN: this component can learn scalable, interpretable patterns from structured features so that it provides a stable baseline and interpretable decision logic-critical for real-world deployment in indoor robotics. The whole architecture is benefited from this DNN, i.e. 1) fast convergence and high accuracy on clean data, 2) easier to visualize and debug, 3) can be regularized (e.g. using different activation function and dropout methods) for generalization. In particular, DNN using CNN-LSTM is regulated by choosing the optimal dropout rate of the number of nodes and layers in the network so that the best performance would be achieved.
    \item Fusion Layer: this aims to combine quantum and classical insights (i.e. DQNN and DNN, respectively) so that the fusion allows the system to adaptively balance robustness (quantum contribution) and interpretability (classical deep leaning contribution) depending on signal quality. We can employ the regular fusion with using simple concatenation and trainable fusion with capability of learning optimal weighting between DQNN and DNN outputs. However, we use the deep learning-based trained fusion to retain the autonomy and accuracy for the framework.
\end{itemize}

Here, our proposed model has the following prominent properties: 1) Cross-validate signals: if one branch is fooled, the other may still detect the anomaly, 2) Adapt to signal quality: fusion can weight branches differently depending on input reliability and 3) Preserve privacy: even with anchor features removed, the hybrid model maintains high Attack F1 Score. 
The first strong contribution is complementary learning, i.e. classical and quantum branches capture orthogonal features, boosting generalization. 
Meaning that when the input is corrupted or spoofed, the system leans more heavily on the DQNN branch, which is designed to handle uncertainty and adversarial noise. 
When the input is clean and normalized, the system favors the DNN branch, which excels at learning structured patterns from stable data.
The other contribution is providing confidence-weighted fusion, which enables dynamic trust assignment based on scenario or sensor reliability.
Besides, our framework has quantum-inspired assignment with interpretability overlays, where parallel outputs allow comparative visualization and error attribution.

Also, we use the small qubit-depth for our study to avoid exponential growth of Hilbert space.
We focused on two tasks:
\begin{itemize}
    \item \textbf{Generalization from Limited Training Data.}  
Fig.~\ref{fig:qnn_generalization} shows the cost function after training and theoretical estimate of the optimal cost function vs number of training pairs.
Our framework closely matches the theoretical bound, demonstrating its strong generalization capability.
    \item \textbf{Robustness to Corrupted Training Data.} 
To assess robustness, we generated $N$ valid training pairs and randomly corrupted $n$ of them by replacing them with random quantum data. 
We then evaluated the cost function on the uncorrupted pairs to measure how well the our model learned the true unitary. 
As shown in Fig.~\ref{fig:qnn_robustness}, our framework exhibits remarkable resilience to such noise, maintaining high fidelity despite data corruption.
\end{itemize}
Furthermore, our Hybrid DQNN+DNN architecture is well-suited to the constraints of NISQ-era hardware. 
The layer-wise structure enables a reduction in the number of coherent qubits required to store intermediate states—scaling only with the network width. 
While estimating gradients requires multiple circuit evaluations, this is a favorable tradeoff given that many NISQ platforms (e.g. superconducting qubits) support rapid circuit repetition. 
The bottleneck in the near term is likely the availability of coherent qubits, and our architecture is designed to operate within this constraint.

In summary, we have introduced natural quantum generalizations of perceptrons and deep neural networks, along with an efficient quantum training algorithm. 
Our Hybrid DQNN+DNN framework demonstrates 1) strong generalization from limited data, and 2) robustness to noisy or corrupted training inputs.

\section{Numerical Results and Discussion on Applications}

\subsection{Data Collection and Preparation}

To evaluate the performance of our proposed method, we use the dataset \cite{pedregosa2011scikit, quigley2009ros, guerrero2017empirical, guerrero2018detection} and then integrate the privacy preservation to regenerate dataset for our testing purpose.
The data collection is briefly summarized as follows.
The robot followed two predefined trajectories—test and validation—under conditions of no attack and DoS attack, where DoS attacks interrupted signals from selected anchors.
The RTLS used six anchors (types A–D) and a mobile tag to estimate 2D positions. 
Each trajectory was repeated 10 times, yielding 20 rosbag files and over 8,400 location estimates.

\subsection{Performance Parameters in Analysis}
\subsubsection{Attack Detection without Privacy Preservation}
The first class-wise performance is the accuracy classification score, which is derived as
\beqn
accuracy = \frac{\sum T_p + \sum T_n}{\sum \mbox{total data}},
\eeqn
where $\sum T_p$ is the number of true positives, and $\sum T_n$ is the number of true negatives.
Besides, the other class-wise performances including Precision ($P$), Recall ($R$) and $F_1$ score are given as
\beqn
P = \frac{\sum T_p}{\sum T_p + \sum F_p},\\
R = \frac{\sum T_p}{\sum T_p + \sum F_n},\\
F_1 = 2\frac{PxR}{P + R},
\eeqn
where $\sum F_p$ and $\sum F_n$ are the number of false positives and number of false negatives, respectively.

\subsubsection{Attack Detection with Privacy Preservation}
In the integration with privacy preserving scheme, we have less data for training so that the performance would decrease. Therefore, there is a need to modify the algorithms such that the detection of attacks must be accurate. To enhance the evaluation step, we also define the additional performances as 
\begin{itemize}
\item $\mbox{Macro }F_1$: Mean of per-class $F_1$s.
\item $\mbox{Weighted }F_1$: Class-prevalence-weighted $F_1$.
\item $\mbox{Attack }F_1$: $F_1$ is specific to attack class. This supports targeted evaluation of DoS detection capability.
\end{itemize}
In the above, we utilize different performances, which are then used to feedback to adjust the parameters of DQNN in the training step such as the dropout rate, the use of DNN or DNN-Shallow.

\subsection{Performance Evaluation and Discussion}

\subsubsection{Performance Comparison Under Privacy Constraints}
Table~\ref{table_main} presents a comparative evaluation of five methods: (1) Fully connected convolutional neural networks (NN); (2) Shallow deep neural networks (DNN-Shallow) with a single hidden layer and limited neurons; (3) Standard deep neural networks (DNN); (4) Hybrid DQNN+NN with 2, 4 and 6 qubits; and (5) Hybrid DQNN+DNN with 2, 4 and 6 qubits.  
To model privacy-preserving conditions in indoor robotic environments, features 4, 5 and 6 are completely removed, and feature 9 is encoded to obfuscate sensitive information, whilst retaining its utility for attack detection.  
This setup reflects realistic constraints, where privacy-aware data sharing is essential.  
We evaluate class-wise performance using five metrics: accuracy, precision, recall, F1-score and training time.

The results show that our proposed hybrid architectures, i.e. Hybrid DQNN+DNN and Hybrid DQNN+NN, consistently outperform traditional models across all metrics, even under feature suppression. 
This highlights the strength of quantum-enhanced processing in extracting complex patterns and high-dimensional correlations from incomplete or obfuscated data.  
In contrast, conventional models (NN, DNN and DNN-Shallow) struggle to recover essential features from the reduced input space, leading to performance saturation near an upper-bound threshold.

\noindent\textbf{Modular Contributions of Our Hybrid Framework}:  
Our architecture integrates three synergistic components:
1) \textbf{DNN-based Dimensionality Reduction in DQNN}: This module compresses the input space, while preserving discriminative features, balancing accuracy and compression ratio.  
It enables efficient feature extraction from
high-dimensional, privacy-filtered data. 
2) \textbf{Quantum Feature Transformation via DQNN}: Quantum gates and circuits simulate complex probability distributions and encode features into quantum states.
This enhances representational capacity and supports robust classification under uncertainty and partial
observability.  
3) \textbf{Classical Pattern Classification via DNN or NN}: The final stage performs pattern classification using
dropout-optimized DNN or NN architectures. 
Regularization of dropping out the layers/nodes in DNN or NN helps tune the network depth and width for optimal detection performance, even under reduced feature availability.

Furthermore, our experiments indicate that the highest classification accuracy is achieved with four qubits for both hybrid architectures. 
This finding suggests that optimal performance can be attained with a relatively small quantum footprint, enabling significant reductions in computational overhead.

Importantly, all quantum experiments were conducted using a local simulator, which supports up to six qubits. 
While this setup enables efficient prototyping, it incurs higher training latency due to limited numerical precision and resource constraints.  
We anticipate that transitioning to IBM Quantum and Qiskit-based simulators will significantly reduce training time, i.e. potentially matching or surpassing DNN training speeds, while further improving detection performance.  
This is due to the superior arithmetic precision, parallel processing capabilities and hardware fidelity of IBM's cloud-accessible quantum infrastructure.  
Note that our proposed Hybrid DQNN+DNN architecture is fully compatible with IBM Quantum's cloud-accessible supercomputing infrastructure. 
Since our experiments utilize only 2, 4 and 6 qubits, the model is well within the operational limits of current NISQ-era devices such as IBM's superconducting qubit platforms. 
This compatibility ensures that transitioning from local simulation to IBM Quantum and Qiskit-based execution is straightforward. 
Leveraging IBM's high-fidelity hardware and parallelized circuit execution is expected to significantly reduce training latency and improve convergence, making our framework readily deployable for real-world CPS security applications.
By minimizing the required number of qubits, while maintaining high accuracy, our framework demonstrates practical scalability for near-term quantum devices and resource-constrained CPS security deployments.

\noindent\textbf{Parallelism Advantage}:  
A key strength of our hybrid framework lies in its parallel architecture, which enables simultaneous execution of quantum and classical branches. 
Unlike serial configurations, where either 1) a DQNN performs feature extraction followed by DNN classification, or 2) a DNN extracts features for subsequent quantum classification, our design avoids cumulative training latency by processing both branches concurrently. 
This parallelism significantly reduces total training time and allows dynamic fusion of quantum robustness with classical interpretability. 
When deployed on IBM's parallelized quantum infrastructure, the overall training time approaches that of the DNN alone, or potentially less, while achieving superior detection performance. 
Compared to traditional models and serial hybrids, our framework offers a more efficient, scalable and explainable solution for real-time CPS security, especially under privacy constraints and adversarial conditions.

\begin{table*}
\centering
\caption{Class-wise performance comparison, when deleting features 4, 5, and 6 completely, while encoding feature 9.}
\label{table_main}
\begin{tabular}{|c|c|c|c|c|c|c|}
\hline 
Model & Qubit Depth & Accuracy & TrainTime & Precision & Recall & F1 Score \tabularnewline
\hline\hline 
DNN & 0 & 0.895 & 40.596 & 0.902 & 0.896 & 0.895 \tabularnewline
\hline 
NN & 0 & 0.795 & 26.899 & 0.832 & 0.796 & 0.790 \tabularnewline
\hline 
DNN-Shallow & 0 & 0.857 & 26.462 & 0.873 & 0.858 & 0.856 \tabularnewline
\hline 
Hybrid DQNN+NN & 2 & 0.908 & 2217.539 & 0.913 & 0.908 & 0.908 \tabularnewline
\hline 
Hybrid DQNN+NN & 4 & 0.910 & 3104.622 & 0.913 & 0.910 & 0.909 \tabularnewline
\hline 
Hybrid DQNN+NN & 6 & 0.900 & 3529.502 & 0.906 & 0.900 & 0.900 \tabularnewline
\hline 
Hybrid DQNN+DNN & 2 & 0.532 & 2271.870 & 0.535 & 0.531 & 0.514 \tabularnewline
\hline 
Hybrid DQNN+DNN & 4 & 0.952 & 3103.029 & 0.952 & 0.952 & 0.952 \tabularnewline
\hline 
Hybrid DQNN+DNN & 6 & 0.929 & 3478.583 & 0.930 & 0.929 & 0.929 \tabularnewline
\hline
\end{tabular}
\end{table*}

\subsubsection{Confusion Matrix Analysis and Safety Implications}
\begin{figure}[!t]
\centerline{\includegraphics[width=0.4\textwidth]{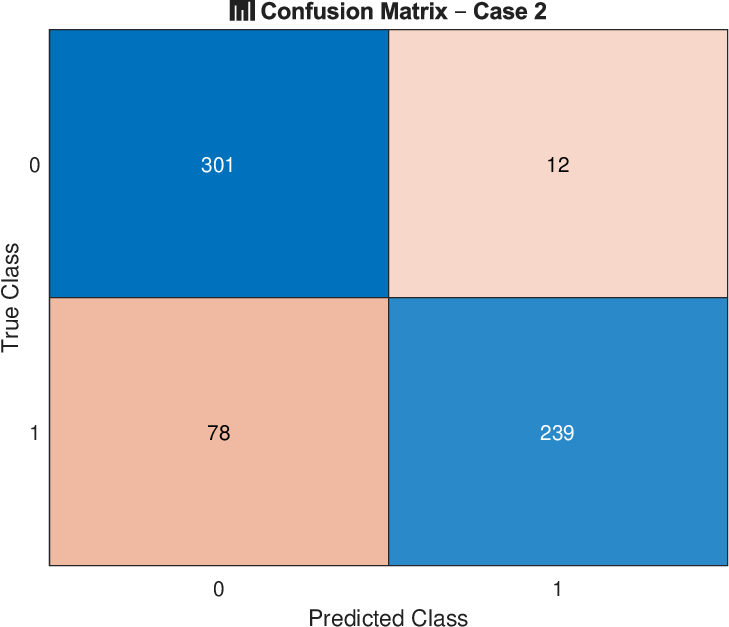}}
\vspace{-0.3cm}
	\caption{Confusion Matrix of DNN, when deleting features 4, 5 and 6 completely, while encoding feature 9.}
	\label{confusion_case2_DNN}
 \vspace{-0.3cm}
\end{figure}

\begin{figure}[!t]
\centerline{\includegraphics[width=0.4\textwidth]{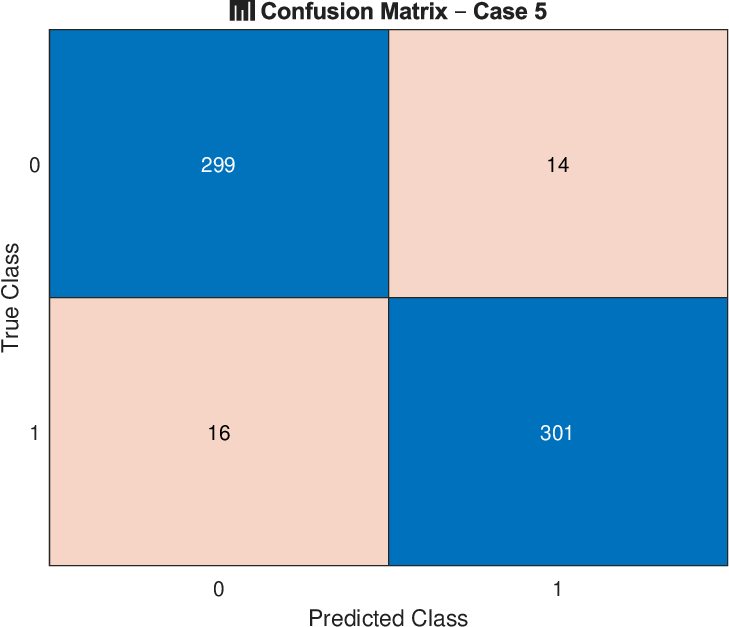}}
\vspace{-0.3cm}
	\caption{Confusion Matrix of Hybrid DQNN and DNN with 4 qubits, when deleting features 4, 5 and 6 completely, while encoding feature 9.}
	\label{confusion_case5_4qubit_HybridDQNN_DNN}
 \vspace{-0.3cm}
\end{figure}

Figs. \ref{confusion_case2_DNN} and \ref{confusion_case5_4qubit_HybridDQNN_DNN} present the confusion matrices for the standard DNN and our proposed Hybrid DQNN+DNN, under the privacy-preserving condition, where features 4, 5 and 6 are completely removed and feature 9 is encoded to obfuscate sensitive information. 
Despite this suppression, the Hybrid DQNN+DNN achieves noticeably higher accuracy than the regular DNN.
More critically, the Hybrid DQNN+DNN demonstrates superior recognition of attack events, significantly reducing false negatives. 
This is essential for robotic networks, where missed detections can lead to cascading failures and collateral damage. 
In such systems, a single undetected malware or spoofing event can compromise localization, control, and coordination, propagating risk across interconnected nodes.
Our proposed method can therefore be deployed within distributed control centers to proactively assess the impact of localized failures. 
By identifying critical nodes, whose compromise could trigger systemic disruption, the framework supports resilience analysis and safety assurance in real-time robotic operations.

\subsubsection{Dropout Rate Optimization and Attack Detection Fidelity}
\begin{figure}[!t]
\centerline{\includegraphics[width=0.55\textwidth]{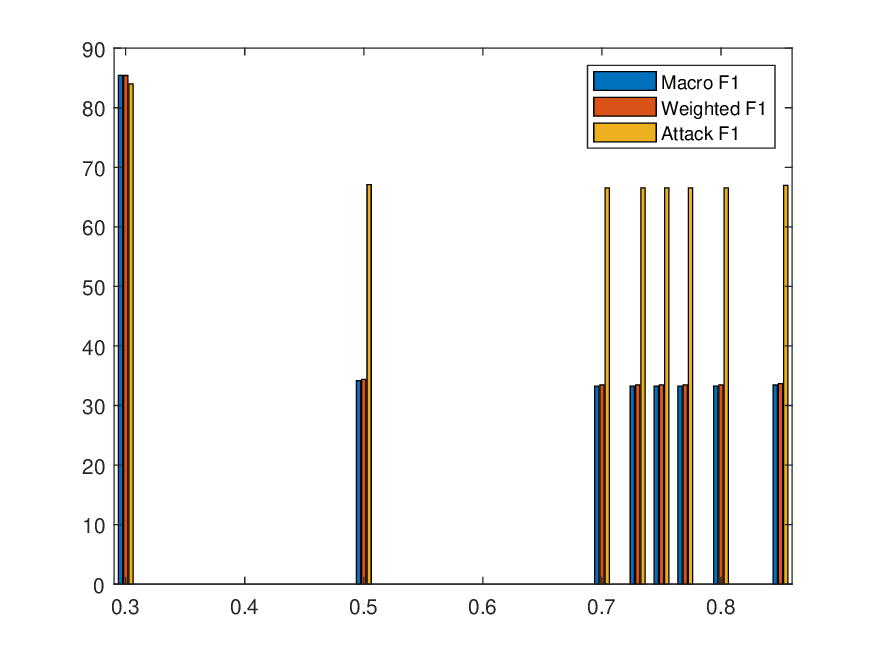}}
\vspace{-0.3cm}
	\caption{NN – Performance vs Dropout Rate, when deleting features 4, 5 and 6 completely, while encoding feature 9.}
	\label{VaF1_vs_dropoutrate_case2}
 \vspace{-0.3cm}
\end{figure}

Fig.~\ref{VaF1_vs_dropoutrate_case2} examines the effect of varying dropout rates on multiple F1 Score metrics, including Macro F1, Weighted F1 and Attack F1 Scores. 
Among these, the Attack F1 Score is particularly vital, as it directly reflects the model's ability to detect malicious events without omission, i.e. an essential requirement for resilient robotic networks.
Our analysis reveals that a dropout rate of approximately 30\% strikes the optimal balance between generalization and precision. 
At this rate, the model maintains high overall accuracy, while preserving its sensitivity to attack-event detection. 
This suggests that moderate regularization not only mitigates overfitting but also enhances robustness against adversarial perturbations, especially when critical features are suppressed or encoded.

\subsubsection{Activation Function and Dropout Rate Optimization}
\begin{figure}[!t]
\centerline{\includegraphics[width=0.4\textwidth]{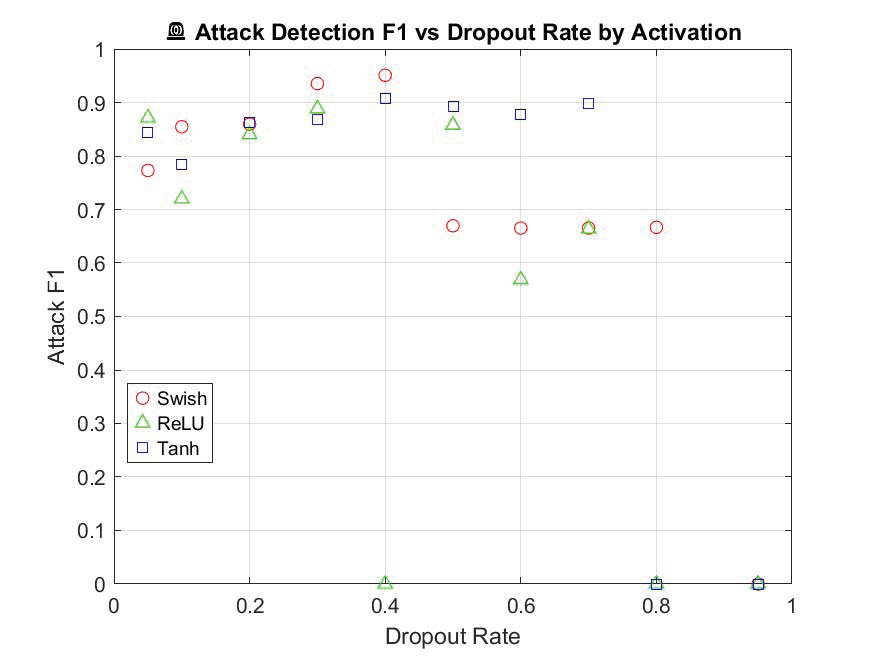}}
\vspace{-0.3cm}
	\caption{Attack detection $F_1$ vs dropout rate by different activation functions (Swish, ReLu and Tanh) for DNN part.}
	\label{Plot_AttackF1_vs_Dropout_AllActivations}
 \vspace{-0.3cm}
\end{figure}

\begin{figure}[!t]
\centerline{\includegraphics[width=0.4\textwidth]{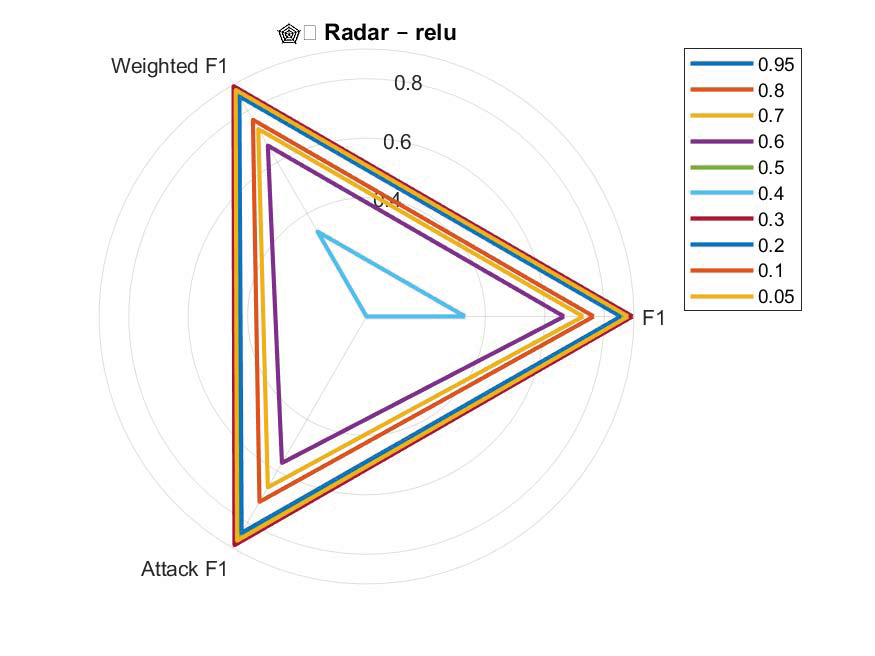}}
\vspace{-0.3cm}
	\caption{Radar representation for different attack detections (Macro $F_1$, Weighted $F_1$ and Acttack  $F_1$) vs dropout rate by ReLu activation function for DNN part.}
	\label{Radar_relu}
 \vspace{-0.3cm}
\end{figure}

\begin{figure}[!t]
\centerline{\includegraphics[width=0.4\textwidth]{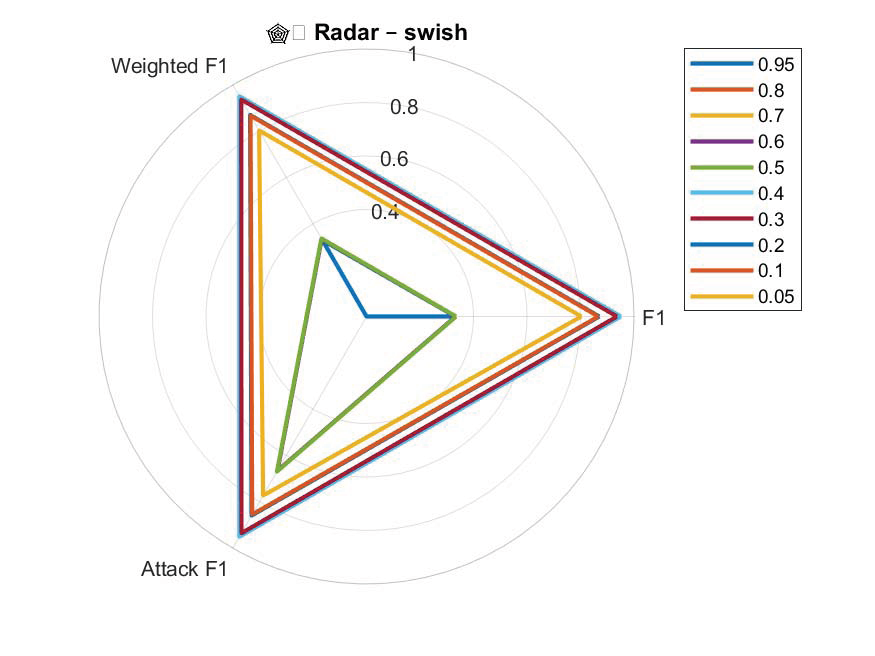}}
\vspace{-0.3cm}
	\caption{Radar representation for different attack detections (Macro $F_1$, Weighted $F_1$ and Acttack  $F_1$) vs dropout rate by Swish activation function for DNN part.}
	\label{Radar_swish}
 \vspace{-0.3cm}
\end{figure}

\begin{figure}[!t]
\centerline{\includegraphics[width=0.4\textwidth]{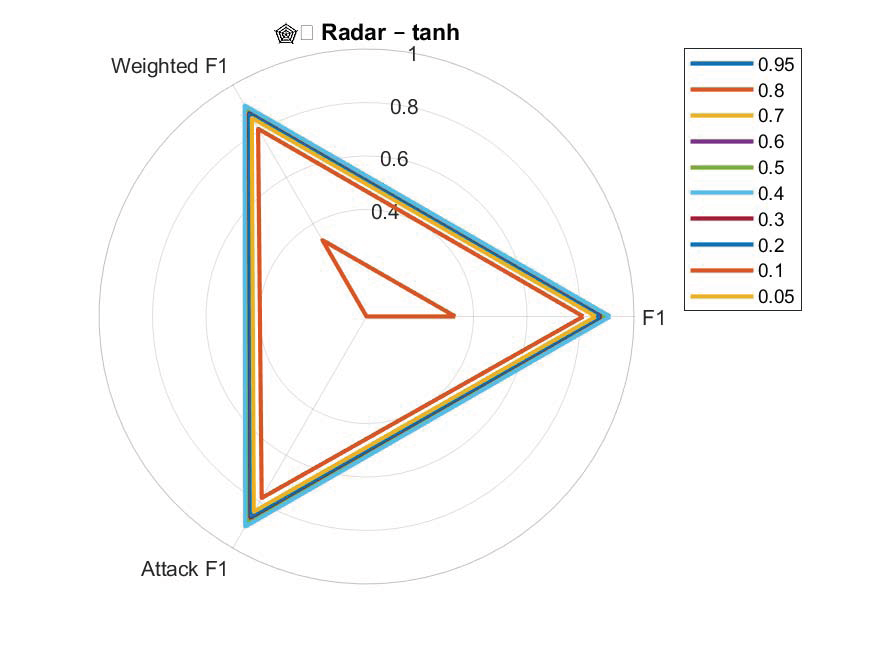}}
\vspace{-0.3cm}
	\caption{Radar representation for different attack detections (Macro $F_1$, Weighted $F_1$ and Acttack  $F_1$) vs dropout rate by Tanh activation function for DNN part.}
	\label{Radar_tanh}
 \vspace{-0.3cm}
\end{figure}

\begin{figure}[!t]
\centerline{\includegraphics[width=0.4\textwidth]{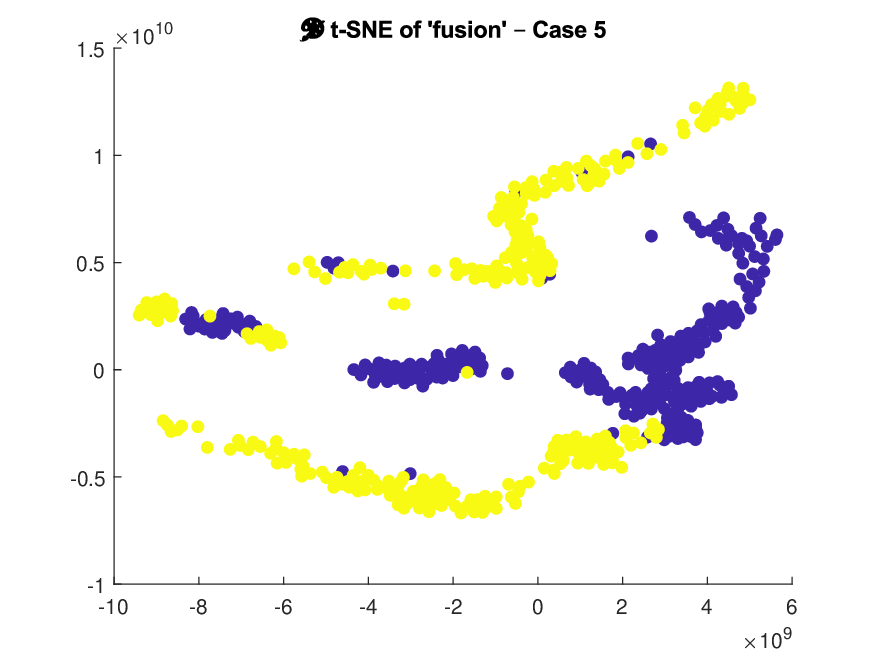}}
\vspace{-0.3cm}
	\caption{t-SNE of Hybrid DQNN and DNN with 4 qubits, when deleting features 4, 5 and 6 completely, while encoding feature 9.}
	\label{tsne_case5_4qubit}
 \vspace{-0.3cm}
\end{figure}

Fig.~\ref{Plot_AttackF1_vs_Dropout_AllActivations} investigates the joint impact of activation function type and dropout rate on the Attack Detection F1 Score within the DNN component of our Hybrid DQNN+DNN architecture. 
Among the tested configurations, the optimal setup is achieved using a dropout rate of 40\% combined with the Swish activation function, which yields the highest and most stable Attack F1 Score.
In contrast, the Tanh activation function performs poorly on this dataset, exhibiting unstable and inconsistent F1 scores across dropout variations. 
This suggests that Tanh may be ill-suited for the high-dimensional, privacy-filtered sharing data used in our malware detection framework.
To provide a clearer comparative visualization, Figs.~\ref{Radar_relu}, \ref{Radar_swish} and \ref{Radar_tanh} present radar plots of Macro F1, Weighted F1, and Attack F1 Scores for the ReLU, Swish and Tanh activation functions, respectively. 
These plots highlight the superior balance and robustness of Swish across all metrics.
Finally, Fig.~\ref{tsne_case5_4qubit} illustrates the structure of the high-dimensional feature space using t-distributed Stochastic Neighbor Embedding (t-SNE). This dimensionality reduction technique preserves local relationships between data points, revealing distinct clusters and latent structure that support effective classification. 
The t-SNE visualization confirms that our hybrid model successfully separates attack and benign instances, even under feature suppression and privacy constraints.

\section{Conclusion and Future Directions}

This work presents a privacy-aware, adversarially robust malware detection framework tailored for indoor robotic systems, leveraging hybrid quantum-classical NNs to counter DoS threats in CPSs. 
By integrating quantum-enhanced feature encoding with deep learning classifiers, the proposed architecture achieves high detection fidelity, interpretability and resilience-even under spoofing, jamming and signal manipulation within intelligent perception systems (IPSs).

Unlike conventional intrusion detection systems, our approach eliminates reliance on handcrafted thresholds or persistent ground-truth beacon data, enabling scalable deployment in dynamic, privacy-sensitive environments. 
The use of VQCs ensures transparent decision-making, while privacy-preserving telemetry analysis protects sensitive robotic data. 
Explainability is further enhanced through confidence-weighted fusion and interpretability overlays, allowing comparative visualization and error attribution across quantum and classical branches.

Benchmark results confirm that our hybrid DQNN+DNN model not only mitigates barren plateau instability but also generalizes effectively across noisy, high-dimensional signal spaces. 
Notably, the optimal configuration, Swish activation with a 40\% dropout rate, yields the highest and most stable Attack F1 Scores, i.e. outperforming traditional activation functions and regularization schemes. 
Furthermore, radar visualizations and t-SNE projections validate the model's robustness and interpretability, even under feature suppression.

Our hybrid DQNN+DNN framework is inherently suited for deployment on current NISQ-era quantum platforms. 
With an optimal configuration requiring only four qubits, the model operates well within the capabilities of devices such as IBM Quantum and IonQ Aria-1. 
Its shallow variational circuits, entanglement-aware design, and efficient gradient estimation make it both hardware-conscious and scalable.
While initial evaluations were conducted using local simulators, the architecture is fully portable to IBM's cloud-based quantum infrastructure. 
This transition is expected to accelerate training, reduce latency and enhance convergence stability, especially under adversarial conditions. 
Moreover, our proposed framework's modular structure and explainability overlays ensure that quantum inference remains interpretable and reproducible, even when executed on remote superconducting backends.

This work contributes to the broader vision of trustworthy AI in robotics by emphasizing reproducibility, modularity, explainability, and ethical impact. 
Future directions include federated quantum learning, real-world IPS integration and adaptive control strategies that dynamically respond to evolving threat landscapes, paving the way for resilient, privacy-preserving autonomy in next-generation robotic platforms.

\bibliography{references}
\bibliographystyle{IEEEtran}

\end{document}